\documentclass[10pt,epsf,letterpaper]{article}

\usepackage{amsmath}
\usepackage{amsfonts}
\usepackage{verbatim}
\usepackage{amssymb}
\usepackage{graphicx}
\usepackage{mathrsfs}
\usepackage{cite}
\usepackage{physics}
\usepackage{hyperref}
\usepackage[utf8x]{inputenc}

\hypersetup{colorlinks=true,	linkcolor=blue,	filecolor=magenta, 	urlcolor=blue,	citecolor=blue,	pdftitle={Holography of AdS Hairy Black Holes and Cardy-Verlinde Formula}, pdfpagemode=FullScreen}

\providecommand{\U}[1]{\protect\rule{.1in}{.1in}}

\def\pa{\partial}

\newcommand{\br}{\biggr}
\newcommand{\bl}{\biggl}

\graphicspath{{./fig_crop/}}
\renewcommand{\(}{\left(}
\renewcommand{\)}{\right)}
\renewcommand{\[}{\left[}
\renewcommand{\]}{\right]}

\begin{document}


\title{Holography of AdS Hairy Black Holes and Cardy-Verlinde Formula}

\author{Dumitru Astefanesei$^{(1)}$, David Choque$^{(1)}$, Jorge Maggiolo$^{(1)}$ and Ra\'ul Rojas$^{(1)}$\\
	\\\textit{$^{(1)}$Pontificia Universidad Cat\'olica de
		Valpara\'\i so, Instituto de F\'\i sica,} \\\textit{Av. Brasil 2950, Valpara\'{\i}so, Chile.} }

\maketitle
\begin{abstract}
We discuss some aspects related to holography of Anti-de Sitter (AdS) dyonic hairy black holes, which break the conformal symetry of the boundary. We use counterterms for the scalar field that satisfies mixed boundary conditions to compute the Euclidean action and dual stress tensor. We apply these results to show that the Cardy-Verlinde formula is not satisfied. However, when the magnetic (or electric) charge vanishes, the conformal symmetry is preserved and the entropy of the black hole can be put in the Cardy-Verlinde form. In our analysis, there is no need of adding extra finite counterterms and, in this particular case, we explicitly show that our results match the ones obtained from using the superpotential as a counterterm.

\end{abstract}
{\hypersetup{linkcolor=black}
\tableofcontents}

\newpage

\section{Introduction}
%
The anti–de Sitter/conformal field theory (AdS-CFT) duality has led to a new method to regularize the Euclidean action of gravity systems \cite{Henningson:1998gx,Skenderis:2000in,Balasubramanian:1999re}. This method is useful for studying the thermodynamics of AdS hairy black holes, e.g. \cite{Batrachenko:2004fd,Skenderis:2002wp,Papadimitriou:2007sj,Anabalon:2015xvl,Anabalon:2017yhv,Anabalon:2020qux,Gnecchi:2014cqa, Astefanesei:2019pfq,Toldo:2016nia} as well as of the compact objects such as boson stars \cite{Astefanesei:2003qy,Astefanesei:2003rw,Buchel:2013uba}, and it was also generalized to asymptotically flat systems \cite{Mann:2005yr,Astefanesei:2005ad,Astefanesei:2009wi,Compere:2011db,Compere:2011ve,Astefanesei:2019mds,Astefanesei:2019qsg}. 

Holographically, the mixed boundary conditions obeyed by the scalar field are associated with a multi-trace deformation of the dual field theory \cite{Witten:2001ua}. Interestingly, the scalar fields can break the conformal symmetry of the AdS boundary \cite{Henneaux:2006hk} such that the trace of the holographic stress tensor of the dual field theory does not vanish. Lu, Pope, and Pang (LPP) have obtained an exact hairy black hole solution with this property in \cite{Lu:2013ura}. However, when the magnetic (or electric) charge vanishes, the conformal symmetry is preserved. We use the counterterms for scalar fields in AdS with mixed boundary conditions proposed in \cite{Anabalon:2015xvl} to compute the Euclidean action and holographic stress tensor for the LPP solution. Since the variational principle is well defined, there is no need of adding extra finite counterterms as in \cite{Liu:2004it}.

An interesting and puzzling (due to its generality, including flat spacetime solutions for which there is no concrete holographic proposal) result of Verlinde \cite{Verlinde:2000wg} is that the entropy of AdS black holes, with a field theory dual, can be put in a form similar to Cardy formula of two-dimensional CFTs. The main observation of Verlinde is that one can define a Casimir energy in the dual theory as the violation of the Euler identity $E_C=2(E+pV-TS)$ (in four bulk dimensions), that is the sub-extensive part of the total energy. For that, one has to consider the  thermodynamic system (in our case, the dual field theory) in a finite volume and the total energy as a function of entropy and volume.\footnote{We emphasize that the analysis presented here is not related in any way with the extended thermodynamics of hairy black holes in \cite{Astefanesei:2019ehu} where the cosmological constant is associated to the pressure.} 

Since in two dimensions the modular symmetry is at the basis of Cardy formula and this symmetry is absent in higher dimensions, it is not clear what is the origin of Cardy-Verlinde formula. However, it was checked for many examples of black holes, e.g. in AdS \cite{Cai:2001jc,Klemm:2001db,Biswas:2003sn}, dS \cite{Cai:2001sn,Jing:2002aq}, and flat spacetime \cite{Youm:2001kw,Jing:2011za}. Particularly, Cai has checked a variant of this formula for a family of hairy charged black holes in AdS \cite{Cai:2001jc}. In this letter, we use holographic tools to put Cai's proposal on a firm ground. We also explicitly show, though, that the Cardy-Verlinde formula is not satisfied for the dyonic LPP hairy black hole solution when the asymptotic conformal symmetry is broken.

This paper is organized as follows. In the next section, the Reissner-Nordstr\"{o}m (RN) black hole in `brane coordinates' and exact hairy black hole solutions relevant to our analysis are briefly presented. In Section \ref{section3}, we use holographic methods to obtain the stress tensor of the dual theory corresponding to hairy black holes. Section \ref{section4}  describes the Cardy-Verlinde formula, establishing that it is always satisfied for the hairy black holes when the conformal symmetry of the AdS boundary is not broken. Finally, we summarize our findings and present some conclusions in the last section.

\section{Exact Hairy Black Hole Solutions}

In this section we present the relevant exact hairy black hole solutions. We show that, up to a redefinition of the solution's parameters, the electrically charged LPP solution is equivalent with a R-charged black hole in a theory with only one gauge field turned on.
\subsection{R-Charged Black Holes}
\label{Rcharged}

We are interested in a truncated version of $D=4$, $N=8$ gauged supergravity and use the conventions of \cite{Gibbons:2005vp}
\begin{equation}
	I[g_{\mu\nu},\vec{\phi}, A_i]=\frac{1}{16\pi}\int_{\mathcal{M}}d^{4}x\sqrt{-g}\bl{[}R- \frac{1}{2}(\pa\vec{\phi})^{2}-\frac{1}{4}\sum_{i}^{4}{X_{i}^{-2}F_i^{2}}+\frac{1}{l^{2}}\sum_{i<j}^{4}X_{i}X_{j}\br{]}
	\label{acti}
\end{equation}
with $G=c=1$. The three scalar fields are written in the compact form $\vec{\phi}=(\phi_{1},\phi_{2},\phi_{3})$; $\{A_i\}$ are the four 1-form gauge potentials, and the four $\{X_{i}\}$, which satisfy the constraint $X_{1}X_{2}X_{3}X_{4}=1$, denote
\begin{equation}
X_{i}=e^{-\frac{1}{2}\vec{a}_{i}\cdot\vec{\phi}}, \qquad \vec{a}_{i}\cdot\vec{a}_{j}=4\delta_{ij}-1
\end{equation}
A convenient choice \cite{Cvetic:1999xp} for the constants $\vec{a}_i$ is
\begin{equation}
	\vec{a}_{1}=(1,1,1), \qquad \vec{a}_{2}=(1,-1,-1), \qquad \vec{a}_{3}=(-1,1,-1), \qquad \vec{a}_{4}=(-1,-1,1)
\end{equation}
with which the action and potential can be rewritten as
\begin{equation}
I[g_{\mu\nu},\vec{\phi},A_i] =\frac{1}{16\pi} \int_{\mathcal{M}}d^{4}x\sqrt{-g} 
\bl{[}R-\frac{1}{2}(\pa\vec{\phi})^{2} -\frac{1}{4} 
\sum_{i}{e^{\vec{a}_{i}\cdot\vec{\phi}}F_i^{2}} -V(\vec{\phi})\br{]}
\end{equation}
\begin{equation}
V(\vec{\phi})=-\frac{2}{l^{2}}\(\cosh{\phi_{1}}+\cosh{\phi_{2}}+\cosh{\phi_{3}}\)
\end{equation}
The equations of motion consistent with this action are
\begin{equation}
	R_{\mu\nu}-\frac{1}{2}g_{\mu\nu}R=T^{\phi}_{\mu\nu}+T^{EM}_{\mu\nu}
\end{equation}
\begin{equation}
	\frac{1}{\sqrt{-g}}\partial_{\mu}(\sqrt{-g}~\partial^{\mu}\vec{\phi})
	=\frac{\pa V(\vec{\phi})}{\pa\vec{\phi}}+\frac{1}{4}\sum_{i}^{4}\frac{\partial (e^{\vec{a}_{i}\cdot\vec{\phi}})}{\partial\vec{\phi}}F_{i}^{~2}
\end{equation}
where the energy-momentum tensors are
\begin{equation}
T^{\phi}_{\mu\nu}=\frac{1}{2} (\pa_{\mu}\vec{\phi})\cdot(\pa_{\nu}\vec{\phi}) -\frac{1}{2}g_{\mu\nu} \[\frac{1}{2}(\pa\vec{\phi})^{2}+V(\vec{\phi})\],\qquad T^{EM}_{\mu\nu}=\frac{1}{2}\sum_{i}^{4}e^{\vec{a}_{i}\cdot\vec{\phi}}\[(F_{\mu\lambda})_{i}(F_{\nu}^{~\lambda})_{i}-\frac{1}{4}g_{\mu\nu}F_{i}^{2}\]
\end{equation}
There exists an exact static charged hairy black hole solution \cite{Duff:1999gh} for this theory. The metric has the following form
\begin{equation}
	ds^{2}=-\bl{(}\prod_{i=1}^{4}{H_{i}}\br{)}^{-1/2}fdt^{2}+
	\bl{(}\prod_{i=1}^{4}{H_{i}}\br{)}^{1/2}(f^{-1}d\rho^{2}+\rho^{2}d\Omega_{2}^{2})
\end{equation}
where the angular line element is $d\Omega_{2}^{2}\equiv d\theta^{2}+\sin^{2}{\theta}d\varphi^{2} $ and the functions $f$ and $H_i$ are
\begin{equation}
f(\rho)=1-\frac{\eta}{\rho} +\frac{\rho^{2}}{l^{2}}\prod_{i=1}^{4}{H_{i}}(\rho), \quad H_{i}(\rho)=1+\frac{q_{i}}{\rho}
\end{equation}
%
The gauge potentials and the quantities $\{X_{i}\}$ are given by
\begin{equation}
	A_{i}=-(1-H_{i}^{-1})\frac{\sqrt{q_{i}(q_{i}+\eta)}}{q_{i}}dt=-\frac{\sqrt{q_{i}(q_{i}+\eta)}}{\rho+q_{i}}dt
	\,, \qquad X_{i}=\bl{(}\prod_{j=1}^{4}H_{j}\br{)}^{1/4}H_{i}^{-1}
\end{equation}
and the scalar fields are
\begin{equation}
	\phi_{1}(\rho)=\frac{1}{2}\ln\bl{(}\frac{H_{1}H_{2}}{H_{3}H_{4}}\br{)}, \quad
	\phi_{2}(\rho)=\frac{1}{2}\ln\bl{(}\frac{H_{1}H_{3}}{H_{2}H_{4}}\br{)}, \quad
	\phi_{3}(\rho)=\frac{1}{2}\ln\bl{(}\frac{H_{1}H_{4}}{H_{2}H_{3}}\br{)}
\end{equation}
%
\subsubsection{RN Black Hole Truncation}
\label{RNbrane}

If we fix $q_{1}=q_{2}=q_{3}=q_{4}\equiv q$, then, the gauge fields become $F_{1}=F_{2}=F_{3}=F_{4} \equiv F$ and the scalar fields vanish, $\phi_{1}=\phi_{2}=\phi_{3} =0$. The action reduces to
\begin{equation}
	\label{truncaRN}
	I=\frac{1}{16\pi}\int_{\mathcal{M}}{d^{4}x\sqrt{-g}}\bl{(}R-F^{2}+\frac{6}{l^{2}}\br{)}
\end{equation}
The metric becomes
\begin{equation}
ds^{2}=-\frac{f}{H^{2}}dt^{2}+H^{2} \(\frac{d\rho^{2}}{f}+\rho^{2}d\Omega_{2}^{2}\), \qquad f(\rho)=1-\frac{\eta}{\rho}+\frac{\rho^{2}}{l^{2}}H(\rho)^{4}, \quad H(\rho)=1+\frac{q}{\rho}
\end{equation}
and the gauge potential is
\begin{equation}
A=-\frac{\sqrt{q(q+\eta)}}{\rho+q}dt
\end{equation}
This is nothing else than RN black hole solution in the so-called `brane coordinates'.

\subsubsection{One-Charge Hairy Black Hole Truncation}
\label{truncacion}

Now, we are interested in the special consistent truncation obtained after making $q_{1}=q\neq 0$ and $q_{2}=q_{3}=q_{4}=0$. In this case, there are three identical scalar fields that can be redefined as $\phi_{1}=\phi_{2}=\phi_{3}\equiv-\frac{\phi(\rho)}{\sqrt{3}}$. Under these considerations, the theory reduces to the action
\begin{equation}
	I=\frac{1}{16\pi}\int_{\mathcal{M}}
	{d^4x\sqrt{-g}
		\left[
		R-\frac{1}{2}(\pa\phi)^2
		-\frac{1}{4}e^{-\sqrt3\phi}F^2
		+\frac{6}{l^2}
		\cosh\left(\frac{\phi}{\sqrt3}\right)
		\right]}
\end{equation}
and the hairy black hole solution is
\begin{equation}
ds^{2}=-\frac{f}{\sqrt{H}}dt^{2} +\sqrt{H} \(\frac{d\rho^{2}}{f} +\rho^{2}d\Omega_2^2\), \qquad A=-\frac{\sqrt{q(q+\eta)}}{\rho+q}dt, \qquad  
\phi=-\frac{\sqrt{3}}{2}\ln{H}
\label{metric11}
\end{equation}
where
\begin{equation}
f(\rho)=1-\frac{\eta}{\rho}+\frac{\rho^2}{l^{2}}\(1+\frac{q}{\rho}\), \qquad H(\rho)=1+\frac{q}{\rho}
\end{equation}

%
\subsection{Dyonic LPP Black Hole}
\label{DyonicSol}
%
In this section we follow \cite{Lu:2013ura}. We start with the action 
\begin{equation}
	I=\frac{1}{16\pi}\int_{\mathcal{M}}
	{d^4x\sqrt{-g}
		\left[
		R-\frac{1}{2}(\pa\phi)^2
		-\frac{1}{4}e^{-\sqrt3\phi}F^2
		+\frac{6}{l^2}
		\cosh\left(\frac{1}{\sqrt3}\phi\right)
		\right]}
	\label{actionfull}
\end{equation}
where $(\pa\phi)^2\equiv g^{\mu\nu}\pa_\mu\pa_\nu\phi$, $F^2\equiv F_{\mu\nu}F^{\mu\nu}$, $F_{\mu\nu}=\pa_\mu A_{\nu}-\pa_\nu A_{\mu}$. 
The exact static dyonic hairy black hole solution was presented in \cite{Lu:2013ura}, and it is
\begin{equation}
ds^2=
-\frac{f\,dt^2}{\sqrt{H_1H_2}}
+\sqrt{H_1 H_2}
\(\frac{d\rho^2}{f}+\rho^2d\Omega_2^2\),
\label{lupopemetric}
\end{equation}
\begin{equation}
A=\frac{\sqrt2(1-\beta_1) f_0}{\sqrt{\beta_1\gamma_2}H_1}\,dt+ \frac{2\sqrt{2\beta_2\gamma_1}\mu}{\gamma_2} \cos\theta\, d\varphi, \qquad \phi =\frac{\sqrt{3}}{2} \ln\left(\frac{H_2}{H_1}\right)
\label{mLPP}
\end{equation}
where
\begin{equation}
\label{horLLP}
f(\rho)=f_0(\rho)+\frac{\rho^2}{l^2}H_1(\rho) H_2(\rho),\qquad f_0(\rho)=1-\frac{2\mu}{\rho}
\end{equation}
and
\begin{equation}
H_{i}(\rho)=\gamma_{i}^{-1}
\[1-2\beta_{i}f_0(\rho)+\beta_1\beta_2f_0(\rho)^2\],\quad
\gamma_{i}\equiv 1-2\beta_{i}+\beta_1\beta_2
\end{equation}
for $i=1,2$. The quantities $\beta_1$, $\beta_2$ and $\mu$ are the integration constants of the solution, related to the mass, and the electric and magnetic charges. 

Notice that, if we fix $\beta_{2}=0$, we recover the one-charge hairy black hole truncation from the Section \ref{truncacion}, that is, the solution (\ref{metric11}), provided the following identification between the parameters
\begin{equation}
\mu=\frac{\eta}{2}, \qquad 
\beta_{1}=\frac{q}{2(\eta+q)}
\end{equation}
%

\section{Holography of Hairy Black Holes}
\label{section3}

In this section, we use the method of \cite{Anabalon:2015xvl} that provides a well defined variational principle and counterterms that regularize the action to obtain the dual stress tensor of AdS hairy black holes presented in the previous section. The advantage of this method is that it can be applied to hairy black holes \cite{Gallerati:2019mzs,Anabalon:2020pez,Gallerati:2021cty} in extended supergravity models when the superpotential is complex. For the electrically charged family, we prove that our results match the ones obtained by using the superpotential as a counterterm \cite{Batrachenko:2004fd}. 

\subsection{Electrically Charged Hairy Black Hole}
\label{ECharged}
%
For the one-charge hairy black hole solution presented in Section \ref{truncacion}, the scalar field has the following fall-off
\begin{equation}
	\phi(r)=\frac{A}{r}+\frac{B}{r^{2}}+O(r^{-3}), \qquad A=-\frac{\sqrt{3}q}{2}, \qquad B=\frac{\sqrt{3}q^{2}}{8}
\end{equation}
where $r=\rho{H}^{\frac{1}{4}}$ is the canonical coordinate. The boundary conditions for the scalar field are given by the function $W(A)$ that satisfies $B=dW(A)/dA$. It is straightforward to verify that $B=\sqrt{3}A^{2}/6$ and $W(A)=\sqrt{3}A^3/18$. Following \cite{Henneaux:2006hk, Anabalon:2015xvl,Hertog:2004dr,Hertog:2004ns}, we point out that this result corresponds to the case in which the conformal symmetry on the boundary is preserved and so the regularized Euclidean action \cite{Anabalon:2015xvl,Marolf:2006nd} is 
\begin{equation}
I^{E}=I_{bulk}^{E}-\frac{1}{8\pi}\int_{\partial\mathcal{M}}d^{3}x\sqrt{h^{E}}K+\frac{1}{8\pi}\int_{\partial \mathcal{M}} d^{3}x\sqrt{h^{E}}\left ( \frac{2}{l}+\frac{l\mathcal{R}}{2} \right)+\frac{1}{16\pi}\int_{\partial\mathcal{M}}{d^{3}x\sqrt{h^{E}}}\bl{(}\frac{\phi^{2}}{2l}+\frac{\sqrt{3}}{18l}\phi^{3}\br{)}
	\label{ac2}
\end{equation}
where
\begin{equation}
	I^{E}_{bulk}=-\frac{1}{16\pi}\int_{\mathcal{M}}
	{d^4x\sqrt{g^{E}}
		\left[
		R-\frac{1}{2}(\pa\phi)^2
		-\frac{1}{4}e^{-\sqrt3\phi}F^2
		+\frac{6}{l^2}
		\cosh\left(\frac{1}{\sqrt3}\phi\right)
		\right]}
	\label{ac1}
\end{equation}
Here, we use a foliation with hypersurfaces $\rho=const$ and the corresponding induced metric $h_{ab}$. The quantities $K$ and $\mathcal{R}$ in Eq. (\ref{ac2}) are the trace of the extrinsic curvature $K_{ab}$ and the Ricci scalar on the boundary, respectively.

For the solution being considered, we have that the energy, electric charge and chemical potential have the following expressions
\begin{equation}
\label{energyconformal}
	E=\frac{1}{2}\(\eta+\frac{q}{2}\),\quad
	Q=\frac{1}{16\pi}\oint_{S^{2}}{e^{-\sqrt{3}\phi}\star F}=\frac{\sqrt{q(q+\eta)}}{4}, \quad \Psi=\frac{4Q}{q+\rho_{+}}
\end{equation}
These quantities are helpful, because, after a lengthy computation of the Euclidean action, we can verify the quantum-statistical relation
\begin{equation}
F=I^{E}T= -(TS+Q\Psi) +\frac{1}{2}\eta + \frac{1}{4}q = E-TS-Q\Psi
\end{equation}
Notice that there is no need of extra finite terms as in \cite{Liu:2004it}.

For completeness, we emphasize that we can also use the superpotential $\mathcal{W}$ as a counterterm \cite{Batrachenko:2004fd} to get the same result. By using this method, the only gravitational counterterm required is
\begin{equation}
	I_{\text{counterterm}}=-\frac{1}{8\pi}\int_{\pa\mathcal{M}}{d^{3}x}\sqrt{-h}\[\mathcal{W}(\phi)+\frac{l\mathcal{R}}{2}\]
\end{equation}
as can be explicitly compared with the asymptotic expansion of the counterterms in (\ref{ac2}).

The quasilocal stress tensor \cite{Brown:1992br} for the action in Eq. (\ref{ac2}), including the boundary term for the scalar field \cite{Anabalon:2015xvl}, is
\begin{equation}
\tau_{ab}\equiv\frac{2}{\sqrt{-h}}\frac{\delta I}{\delta h^{ab}} =-\frac{1}{8\pi}\[K_{ab}-\(K-\frac{2}{l}\)h_{ab} -lG_{ab}\] -\frac{h_{ab}}{16\pi} \[\frac{\phi^{2}}{2l}+\frac{W(A)}{lA^{3}}\phi^{3}\]
\end{equation}
where $\mathcal{R}_{ab}$ is the Ricci tensor on the boundary and $G_{ab}=\mathcal{R}_{ab}-\frac{1}{2}h_{ab}\mathcal{R}$. The regularized dual stress tensor is related to the quasilocal stress tensor by the conformal transformation \cite{Myers:1999psa}
\begin{equation}
\label{dualst}
\langle \tau_{ab}^{dual} \rangle= \lim_{\rho\rightarrow \infty} \frac{\rho}{l}\tau_{ab}
\end{equation}
 and its components are
\begin{equation}
\langle \tau_{tt}^{dual} \rangle =\frac{1}{8\pi l^{2}}\(\eta+\frac{q}{2}\)=\frac{E}{4\pi l^2}, \qquad 	\langle \tau_{\theta\theta}^{dual} \rangle =\frac{E}{8\pi}, \qquad \langle	\tau_{\phi\phi}^{dual}  \rangle =\sin^{2}\theta	\langle \tau_{\theta\theta}^{dual} \rangle
\end{equation}
This stress tensor is covariantly conserved and its trace vanishes, $\langle \tau^{dual} \rangle =0$, as expected from the fact that the boundary conditions preserve the conformal symmetry.
%
\subsection{Dyonic Hairy Black Hole}
%
\label{dyonicLPP}

For the LPP solution from Section \ref{DyonicSol}, the scalar field falls-off as $\phi(r)=\frac{A}{r}+\frac{B}{r^2}+O(r^{-3})$, where
\begin{equation}
A=\frac
{2\sqrt{3}(\beta_2-\beta_1)(1-\beta_1\beta_2)}
{\gamma_1\gamma_2}\,\mu, \qquad B=\frac{2\sqrt{3}(\beta_2-\beta_1)\[4\beta_1\beta_2 (\gamma_1+\gamma_2)-\(1 -\beta_1\beta_2\)^2(\beta_1+\beta_2)\]}{\gamma_1\gamma_2}\,\mu^2
	\label{fall1}
\end{equation}
In general, these boundary conditions for the scalar field do not preserve the isometries of AdS at the boundary \cite{Henneaux:2006hk, Anabalon:2015xvl}, unless $\beta_1=0$ or $\beta_2=0$. Since the conformal symmetry is broken, there are subtleties with a correct definition of the black hole energy that are discussed in great detail in \cite{Hertog:2004ns,Anabalon:2014fla, Lu:2014maa,Astefanesei:2018vga} and we do not repeat those issues here.

We are going to keep the discussion general and obtain the dual stress tensor as a function of $W$ that will be sufficient for the next section. Following \cite{Anabalon:2015xvl}, the counterterm for the scalar field with mixed boundary conditions is
\begin{equation}
	I^{E}_{\phi}=\frac{1}{16\pi}\int{d^{3}x\sqrt{h^{E}}}\bl{[}\frac{\phi^{2}}{2l}+\frac{W(A)}{lA^{3}}\phi^{3}\br{]}
	\label{contri}
\end{equation}

The electric and magnetic conserved charges are
\begin{equation}
	Q=\frac{\mu\sqrt{\beta_{1}\gamma_{2}}}{\gamma_{1}\sqrt{2}}, \quad P=\frac{\mu\sqrt{\beta_{2}\gamma_{1}}}{\gamma_{2}\sqrt{2}}
\end{equation}
and their conjugate electric and magnetic potentials have the following expressions:
\begin{align}
	\Psi&=\sqrt{\frac{2}{\beta_1\gamma_2}}
	\left[
	1-\beta_1-\frac{1-\beta_1f_0(\rho_+)}{H_1(\rho_+)}
	\right], \\ \Upsilon&=\sqrt{\frac{2}{\beta_2\gamma_1}}
	\left[
	1-\beta_2-\frac{1-\beta_2f_0(\rho_+)}{H_2(\rho_+)}
	\right]
\end{align}
where $\rho_+$ is the coordinate of the outer horizon of the black hole, $f(\rho_+)=0$, from Eq. (\ref{horLLP}).
These expressions are going to be helpful for the following results.

The components of the regularized dual stress tensor and its non-vanishing trace are
\begin{equation}
\label{tautt}
\langle \tau_{tt}^{dual} \rangle =\frac{M}{4\pi l^{2}}+\frac{1}{16\pi l^{4}%
}\biggl{(}W-\frac{AB}{3}\biggr{)}, \quad \langle \tau_{\theta \theta}^{dual} \rangle =\frac{M}{8\pi}-\frac
{1}{16\pi l^{2}}\biggl{(}W-\frac{AB}{3}\biggr{)}, \quad \langle \tau_{\phi \phi}^{dual} \rangle = \sin^{2}\theta \langle \tau_{\theta \theta}^{dual} \rangle 
\end{equation}
\begin{equation}
 \langle \tau^{dual} \rangle = -\frac{3}{16\pi l^{4}}\left ( W-\frac{AB}{3} \right )
\end{equation}
In these expressions, we have used $M$ as a new parameter
\begin{equation}
M=\frac{1}{2}\(\eta+\frac{q}{2}\)
\end{equation}
By computing the Euclidean action, including the counterterm in Eq. (\ref{contri}), we obtain the quantum-statistical relation
\begin{equation}
F=I^{E}T=E-TS-Q\,\Psi-P\,\Upsilon
\end{equation}
where
\begin{equation}
    E=M+\frac{1}{4l^{2}}\bl{(}W-\frac{AB}{3}\br{)}
\end{equation}
is conserved energy $E$ \cite{Brown:1992br}, read from the $\tau_{tt}$ component of the dual stress tensor (\ref{tautt}).

In general, the energy receives a correction from the scalar field. One can explicitly check that, if $B \sim A^{2}$, the trace of the stress tensor $\langle \tau^{dual} \rangle =0$, the conformal symmetry is restored and $E=M$.
%
\section{Cardy-Verlinde formula}
\label{section4}
In this section, we use the previous holographic results to put on a firm ground the Cai's proposal \cite{Cai:2001jc} for a consistent Cardy-Verlinde formula for charged hairy black holes.  

For clarity, we briefly present the Cardy-Verlinde formula and define the relevant physical quantities. First, to simplify the notation, even if the dual stress tensor is an operator, we are going to denominate it as $\tau_{ab}^{dual}$ instead of $\langle\tau_{ab}^{dual}\rangle$. What we have computed is the stress tensor of the dual CFT in the UV regime, namely, at the boundary of AdS. However, 
in the context of Cardy-Verlinde formula, the CFT is living  on a sphere of finite radius $R$. Consequently, a conformal transformation 
\begin{equation}
\gamma_{ab}dx^{a}dx^{b}=\frac{R^{2}}{l^{2}}\[-dt^{2}+l^{2}(d\theta^{2}+\sin^{2}{\theta}d\varphi^{2})\]
\label{met1}
\end{equation}
together with a redefinition of the time coordinate $t=\frac{l}{R}\tau$, put the metric in the required form for the Cardy-Verlinde (CV) formula,
\begin{equation}
\gamma^{CV}_{ab}dx^{\prime a}dx^{\prime b}=-d\tau^{2}+R^{2}(d\theta^{2}+\sin^{2}{\theta}d\varphi^{2})
\label{CVspace}
\end{equation}

Once the holographic stress tensor compatible with this metric is computed, we obtain the pressure and energy density of the dual field theory and verify explicitly the Cardy-Verlinde formula \cite{Verlinde:2000wg}
\begin{equation}
S=\pi R\sqrt{E_{C}(2E-E_{C})}, \qquad E_{C}=2(E+pV-TS)
\end{equation}
where $E_C$ is the Casimir energy, and its extension for charged hairy black holes.

\subsection{RN Black Hole}

As an warm-up exercise, let us first consider the RN black hole in AdS spacetime. The Cardy-Verlinde formula is still valid, but with a small change in the energy expression, as it was presented in \cite{Cai:2001jc}. The novelty of this section is that we provide a physical interpretation of this result, by relating it to Ruffini's irreducible mass \cite{Ruffini:2002rk}.

The RN black hole solution in canonical coordinates is
\begin{equation}
	ds^{2}=-f(r)dt^{2}+f(r)^{-1}dr^{2}+r^{2}(d\theta^{2}+\sin^{2}{\theta}d\varphi^{2}), \qquad f(r)=1-\frac{2M}{r}+\frac{Q^{2}}{r^{2}}+\frac{r^{2}}{l^{2}}
\end{equation}
Now, as a consequence of having changed $t\rightarrow \frac{l}{R}\tau$, we have that some thermodynamic quantities require a corresponding rescaling. The thermodynamic quantities for this solution are
\begin{equation}
	E=\frac{l}{R}M=\frac{l}{2R}\bl{(}r_{+}+\frac{r_{+}^{3}}{l^{2}}+\frac{Q^{2}}{r_{+}}\br{)}, \qquad \Psi=\frac{Ql}{Rr_{+}}, \qquad S=\pi r_{+}^{2}, \qquad T=\frac{l}{4R\pi}\bl{(}\frac{3r_{+}}{l^{2}}+\frac{1}{r_{+}}-\frac{Q^2}{r_{+}^{3}}\br{)}
	\label{RNcant}
\end{equation}
where $r_+$ is coordinate of the outer horizon, $f(r_+)=0$. The thermodynamic volume of the system is $V=4\pi R^{2}$ and the first law of the dual field theory is 
\begin{equation}
dE=TdS-pdV+\Psi dQ, \qquad 	p\equiv -\bl{(}\frac{\pa{E}}{\pa V}\br{)}_{Q,S} =-\(\frac{\pa{E}}{\pa R}\)_{Q,r_+} \(\frac{\pa{R}}{\pa V}\)=\frac{E}{2V}
\end{equation}

It was observed in \cite{Cai:2001jc} that a modified Cardy-Verlinde is valid in this case, namely,
\begin{equation}
	S=\pi R\sqrt{E_{C}[2(E-E_{Q})-E_{C}]}=\pi r_{+}^{2}
\end{equation}
where 
\begin{equation}
E_{C}=2(E+pV-TS-Q\Psi)=\frac{lr_{+}}{R}\,, \quad E_{Q}=\frac{\Psi Q}{2}={\frac {l{Q}^{2}}{2Rr_{+}}}
\end{equation}
Interestingly, $E_Q$ is nothing else than Ruffini's energy of the electromagnetic field that can be extracted from the black hole \cite{Ruffini:2002rk}, that is,
\begin{equation}
E_{Q}\equiv\frac{1}{8\pi}\int_{\Sigma_{t}}{\xi^{\mu}T_{\mu\nu}^{EM}d\Sigma^{\nu}}=\frac{1}{8\pi}\int d^{3}x\sqrt{h}\,\xi^{\mu}n^{\nu}\,T_{\mu\nu}^{EM} = -\frac{Q^{2}}{2r}\br{\vert}_{r_{+}}^{\infty} = {\frac {{Q}^{2}}{2r_{+}}}
\end{equation}
where $n^{\mu}$ is the normal on the surface $\Sigma_{t}$ (defined by $t=const.$), $\xi_{\mu}=\partial/\partial t$ is the Killing vector and 
\begin{equation}
\sqrt{h}=\frac{r^{2}\sin^{2}\theta}{\sqrt{f(r)}}\,, \quad T_{00}=\frac{f(r)Q^{2}}{r^{4}}
\end{equation}
This definition for $E_Q$ \cite{Ruffini:2002rk} is rather in the spirit of the definition of the quasilocal charges \cite{Brown:1992br} of Brown and York with respect to the corresponding Killing vector and it is not the purely electromagnetic energy obtained by the usual integration of the energy density outside the horizon,
\begin{equation}
\int_{\Sigma_t} n^\mu\, T_{\mu\nu}^{EM}d\Sigma^\nu = \frac{1}{2} \int_{r_+}^{\infty} \sqrt{g_{rr}} \frac{Q^2}{r^2} dr
\end{equation}
%
\subsection{Cardy-Verlinde formula for Hairy Black Holes}

The scalar field that represents degrees of freedom living outside the horizon is going to also get subtracted from the total energy that enters in Cardy-Verlinde formula. However, a direct computation -as in the case of RN black hole- produces a divergent result and, at this point, it is not clear to us how it can be regularized. We are going to consider an alternative approach proposed by Cai \cite{Cai:2001jc} and provide a holographic interpretation of this proposal using the dual stress tensor to get the pressure and energy density.

The hairy black hole metric (\ref{metric11}) is written in the so called `brane coordinate' system. The interesting observation made by Cai in \cite{Cai:2001jc} is that, to get a consistent Cardy-Verlinde formula for hairy black holes, one should use a proper internal energy obtained by subtracting the mass of supersymmetric background from the total energy. 

For the hairy electrically charged solution from Section \ref{truncacion}, the rescaled thermodynamic quantities (obviously, not the entropy and electric charge) are 
\begin{equation}
E=\frac{l}{2R}\(\eta+\frac{q}{2}\) ,\;\;
T=\frac{1}{4R}\frac{3\rho_{+}^{2}+2q\rho_{+}+l^{2}}{\pi l\sqrt{\rho_{+}(\rho_{+}+q)}} ,\;\;
S=\pi \rho_{+}^{2}\sqrt{1+\frac{q}{\rho_{+}}} ,\;\;
Q=\frac{\sqrt{q(q+\eta)}}{4},\;\; \Psi=\frac{4Ql}{(q+\rho_{+})R}
\end{equation}

Following the method of Cai, we can identify the energy $E_{q}$ by doing $\eta=0$,
\begin{equation}
E_q=\lim_{\eta\rightarrow 0}{E}
=\frac{ql}{4R}
\end{equation}
and the pressure is obtained by the expression
\begin{equation}
\bar{p}\equiv\frac{E-E_{q}}{2V}  =\frac{\eta l}{16\pi{R}^3}
\end{equation}
The Casimir energy, $\bar{E}_{C}$, is computed as usual, though we also use the pressure from which the background contribution was extracted, that is,
\begin{equation}
	\bar{E}_{C}=2(E+\overline{p}V-TS-\Psi Q)=\frac{l\rho_{+}}{R}
\end{equation}
and the entropy, then, can be put in the following form
\begin{equation}
\label{CVhairyCAI}
	S=\pi R\sqrt{\bar{E}_{C}[2(E-E_{q})-\bar{E}_{C}]}
\end{equation}
which is the Cardy-Verlinde formula for hairy black holes.

Consider now the general metric that describes the dyonic black hole (\ref{lupopemetric}) and the rescaled holographic stress tensor that corresponds to a theory associated with the $(2+1)$-dimensional metric, $\gamma^{CV}_{ab}$, (\ref{CVspace}):
\begin{equation}
	\tau^{CV}_{\tau\tau}=\frac{l}{4\pi R^{3}}\biggl{[}M+\frac{1}{4l^{2}%
	}\biggl{(}W-\frac{AB}{3}\biggr{)}\biggr{]}, \quad 
	\frac{\tau_{\varphi\varphi}^{CV}}{\sin^{2}\theta}=\tau_{\theta\theta}^{CV}=
	\frac{l}{8\pi R}\biggl{[}M-\frac
	{1}{2l^{2}}\biggl{(}W-\frac{AB}{3}\biggr{)}\biggr{]}
\end{equation}
As we have already shown in Section \ref{dyonicLPP}, for general mixed boundary conditions for the scalar field, the dyonic LPP black hole does not preserve the isometries of AdS at the boundary. Putting the stress tensor in the form of a thermal gas,
\begin{equation}
	\tau_{ab}^{CV}=(\epsilon+p)u_{a}u_{b}+p\gamma_{ab}
	\label{space3}
\end{equation}
we can read the pressure and energy density  
\begin{equation}
	\epsilon=\frac{1}{V}~\frac{l}{R}\biggl{[}M+\frac{1}{4l^{2}%
	}\biggl{(}W-\frac{AB}{3}\biggr{)}\biggr{]}, \qquad
	p=\frac{1}{2V}~\frac{l}{R}\biggl{[}M-\frac
	{1}{2l^{2}}\biggl{(}W-\frac{AB}{3}\biggr{)}\biggr{]}
\end{equation}
where, again, the volume is $V=4\pi R^{2}$. For a general boundary condition, the trace of the stress tensor does not vanish and the energy receives a correction from the scalar field. As expected, the Cardy-Verlinde formula is not valid in this case.

However, when the magnetic (or the electric) charge vanishes, e.g. $\beta_{2}=0$ ($\beta_1=0$) as in Section (\ref{ECharged}), we have that $W-\frac{AB}{3}=0$ and now the trace of the stress tensor vanishes. This explicitly proves that, for this particular case, the conformal symmetry in the boundary is recovered, the pressure and energy density become 
\begin{equation}
	\epsilon=\tau_{\tau\tau}^{CV}=\frac{l}{2V  R}\left (\eta + \frac{q}{2} \right), \qquad ~  p=\frac{\tau_{\theta\theta}^{CV}}{R^{2}}= \frac{l}{4VR}\left ( \eta + \frac{q}{2} \right)
\end{equation}
and so we obtain the stress tensor of a perfect gas of massless particles with energy and pressure
\begin{equation}
\label{Eq}
E=\frac{l}{2R}\left ( \eta+\frac{q}{2} \right), \qquad
p=\frac{E}{2V}
\end{equation}
Now, we have all the holographic ingredients to check Cai's proposal. Under this consideration, we can separate the energy in three different parts
\begin{eqnarray}
E=E_{E}+\frac{1}{2}\bar{E}_{C}+E_{q}, \qquad E_{q}=\frac{ql}{4R}, \qquad E_{E}=E-E_q-\frac{1}{2}\bar E_C =\frac{1}{2}\frac{\rho_{+}^{2}(\rho_{+}+q)}{lR}
\end{eqnarray}
As in the original work of Verlinde \cite{Verlinde:2000wg}, we interpret the term $E_E$ as the purely extensive part of the energy and $\bar{E}_{C}$ as the Casimir energy (the sub-extensive part). Additionally, we have that $E_q$  is the energy of the supersymmetric background when $\eta=0$, as proposed by Cai in \cite{Cai:2001jc}. A straightforward computation proves that the Cardy-Verlinde formula (\ref{CVhairyCAI}) is satisfied for hairy black hole solutions that are asymptotically locally AdS.

\section{Conclusions}
We have analyzed the holography of dyonic LPP AdS black hole for which the conformal symmetry on the boundary is broken. We have shown that, up to a redefinition of black hole parameters, the electrically charged LPP solution (when the magnetic charge vanishes) matches an R-charged black hole truncation with only one gauge field turned on for which the conformal symmetry of the boundary is preserved. Therefore, we have a nice set-up where to explicitly verify some holographic properties of hairy black holes. Particularly, we have checked the validity of Cardy-Verlinde formula in this case. When the conformal symmetry is preserved, the entropy of hairy black hole can be put in the Cardy-Verlinde form, otherwise this result is spoiled. This should not come as a surprise since, as we have explicitly shown,  the energy has an extra contribution from the scalar field when the trace of the dual stress tensor does not vanish. This may hint that, for the asymptotically flat black holes for which Cardy-Verlinde formula is also valid (and for which there is no concrete holographic proposal), there exists a hidden conformal symmetry. A similar proposal exists for explaining the universality of black hole entropy, e.g. \cite{Carlip:2007qh, Carlip:1998wz}. 

In \cite{Cai:2001jc}, it was shown that a variant of this formula is also satisfied for hairy black holes, particularly the R-charged black holes. In Section \ref{section4}, we have provided a holographic computation of the stress tensor that supports this proposal. In the case of RN black hole, we also gave a  physical interpretation of the energy used in Cardy-Verlinde formula by relating it to the irreducible mass of Ruffini \cite{Ruffini:2002rk}. In the presence of the scalar field, we did not succeed to compute the regularized contribution to be able to compare the two relevant quantities. However, we have used an alternative method put forward by Cai in \cite{Cai:2001jc}. In the `brane coordinate' system, the relevant contributions to the energy can be separated and it is possible to subtract the mass of the (supersymmetric) background with $\eta=0$ (\ref{Eq}). For the RN black hole presented in Section \ref{RNbrane}, the solution with $\eta=0$ is the extremal black hole and one has to do a background subtraction with respect to the zero temperature state. One obvious problem with this proposal, when applied to hairy black holes, is that the background is a naked singularity, but we know that in string theory this solution can be interpreted as a system of giant gravitons \cite{Myers:2001aq} and so everything fits nicely in this context.

\section{Acknowledgments}
The work of DA and RR was supported by the Fondecyt grant 1200986. DA is further supported by Proyecto de Cooperaci\'on Internacional 2019/13231-7 FAPESP/ANID.~ DC would like to thank to PUCV and Univ. Micaela Bastidas for the hospitality during the stages of this research.~JM studies are supported by
CONICYT (currently ANID) grant No. 21212072.

\end{document}